\documentclass[utf8]{FrontiersinHarvard} 

\usepackage{url,lineno,microtype,subcaption}
\usepackage[onehalfspacing]{setspace}
\usepackage{amsmath,amsfonts}
\usepackage{adjustbox}
\usepackage{float}
\usepackage{textcomp}
\usepackage{stfloats}
\usepackage{url}
\usepackage{verbatim}
\usepackage{graphicx}
\usepackage{booktabs}
\usepackage{tikz}
\usetikzlibrary{shapes.geometric,positioning}
\usepackage{color}
\usepackage{lmodern}
\usepackage{eurosym}
\usepackage{hyperref}
\usepackage{nomencl}
\makenomenclature
\def\keyFont{\fontsize{8}{11}\helveticabold }
\def\firstAuthorLast{Alejandro Martín-Crespo {et~al.}} 
\def\Authors{Alejandro Martín-Crespo\,$^{1,*}$, Alejandro Hernández-Serrano\,$^{1}$, Óscar Izquierdo-Monge\,$^{2}$, Paula Peña-Carro\,$^{2}$, Ángel Hernández-Jiménez\,$^{2}$, Fernando Frechoso-Escudero\,$^{3}$ and Enrique Baeyens\,$^{4}$}
\def\Address{$^{1}$Centro Tecnológico CARTIF, Boecillo, Spain \\
$^{2}$Centro de Desarrollo de Energías Renovables (CEDER)—Centro de Investigaciones Energéticas Medioambientales y Técnológicas (CIEMAT), Soria, Spain \\
$^{3}$Departamento de Ingeniería Eléctrica, Escuela de Ingenierías Industriales, Universidad de Valladolid, Valladolid, Spain \\
$^{4}$Instituto de las Tecnologías Avanzadas de la Producción, Universidad de Valladolid, Valladolid, Spain 
}
\def\corrAuthor{Alejandro Martín-Crespo}

\def\corrEmail{alemar@cartif.es}

\begin{document}
\onecolumn
\firstpage{1}

\title[AC/DC OPF and TEA for microgrids: TIGON CEDER]{AC/DC optimal power flow and techno-economic assessment for hybrid microgrids: TIGON CEDER demonstrator} 

\author[\firstAuthorLast ]{\Authors} 
\address{} 
\correspondance{} 

\extraAuth{}

\maketitle

\begin{abstract}

\section{}
In the recent years, the interest in electric direct current (DC) technologies (such as converters, batteries, electric vehicles, etc.) is increasing due to its potential on energy efficiency and sustainability. However, the vast majority of electric systems and networks are based on alternating current (AC), as they also have certain advantages regarding cost-effective transport and robustness. In this paper, an AC/DC optimal power flow method for hybrid microgrids and several key performance indicators (KPIs) for its techno-economic assessment are presented. The combination of both calculations allows users to clearly determine the viability of their hybrid microgrids. AC/DC networks have been modelled considering their most common elements. For the power flow method, a polynomial optimisation is formulated considering four different objective functions: the minimisation of energy losses, voltage deviation and operational costs, and also the maximisation of the microgrid generation. The power flow method and the techno-economic analysis have been implemented in Python and validated in the Centro de Desarrollo de Energ\'ias Renovables (CEDER) demonstrator for TIGON. The results show that the calculated power flow variables and the ones measured at CEDER are practically the same. In addition, the KPIs have been obtained and compared for four operating scenarios: baseline, no battery, battery flexibility and virtual battery (VB) flexibility. The last one result in the most profitable option.

\tiny
 \keyFont{ \section{Keywords:} AC/DC optimal power flow, hybrid microgrids, KPIs, techno-economic assessment, polynomial optimisation, Python} 
\end{abstract}

\section{Introduction}

The global shift towards decarbonization has propelled significant transformations in the design, operation, and management of electric grids. The urgent need to mitigate climate change has led to the adoption of renewable energy sources and the phasing out of fossil fuel-based power generation, resulting in a paradigm shift in the electricity sector.
The integration of variable renewable sources, such as solar and wind, presents unique challenges due to their intermittent nature and geographical distribution. As a result, electric grids have witnessed a remarkable transition towards more dynamic, flexible, and intelligent systems.

Microgrids \citep{lasseter2002microgrids,hatziargyriou2007microgrids,katiraei2008microgrids,salam2008technical,saeed2021review} are localized and self-contained electricity distribution systems. They have gained prominence due to their ability to effectively integrate distributed energy resources (DERs). DERs include small-scale renewable energy installations, energy storage systems, and demand response capabilities. Microgrids provide an innovative solution to enhance the resilience, reliability, and sustainability of the electric grid at a smaller scale, while offering opportunities for local energy generation, utilization, and management. In general, microgrids can work as an island or connected to the main power network which acts as an external grid \citep{marnay2015microgrid}.

Furthermore, the choice between alternating current (AC) and direct current (DC) elements within microgrids has become a subject of considerable interest \citep{wang2013harmonizing}. While AC has historically been the dominant standard for power transmission and distribution, recent advancements in DC technologies, such as state-of-the-art batteries and electric vehicles (EVs), have brought attention to these systems \citep{shao2010review,shi2017constant,fotopoulou2021state}.

In microgrids, where local generation and consumption are tightly integrated, DC elements offer several benefits \citep{saeedifard2010dc,zubieta2016microgrids,rauf2016application,pires2023dc}. Firstly, DC distribution systems enable higher efficiency in the utilization of renewable energy sources. Most renewable energy technologies, such as solar panels and batteries, inherently generate and store DC power. By directly integrating these DC sources into the microgrid without the need for AC-DC conversions, energy losses associated with multiple conversions can be minimized, resulting in improved overall system efficiency. Moreover, DC systems offer increased flexibility for the integration of emerging technologies. As the demand for EVs grows, DC charging infrastructure becomes crucial. DC microgrids can seamlessly accommodate EV charging stations without the need for additional power conversion equipment, reducing infrastructure costs and improving charging efficiency \citep{ashique2017integrated}.

Despite that, AC technologies still offer certain advantages that make them relevant and preferred in specific aspects of microgrids design and operation. For instance, high voltage AC transmission systems inherently offer better voltage control than DC through reactive power, they are easier to isolate and interrupt in case of faults, and the existing infrastructure is more abundant. In the end, both AC and DC technologies have their unique advantages and limitations, and their selection should be based on careful evaluation and analysis of the specific circumstances. In this context, hybrid AC/DC microgrids emerge as a suitable solution for the transition to an electricity system with reduced or zero greenhouse gases emissions, taking advantage of the benefits of both forms of electricity current. One example of a project that seeks to maximise the benefits of these networks is the Horizon 2020 European project called Towards Intelligent DC-based hybrid Grids Optimizing the network performance (TIGON). It also aims to improve reliability, resilience, performance and cost efficiency of hybrid AC/DC grids. 

The main contribution of this article is the development of a procedure to study and evaluate the correct operation and the technical and economic feasibility of hybrid microgrid installations. The developed procedure consists of two components. The first component is a power flow calculation method for hybrid AC/DC microgrids based on optimization. The power flow can be performed by choosing among four different cost functions, depending on the objective to be achieved.
The second component is a techno-economic evaluation based on key performance indicators (KPIs). Another contribution of the article is the validation of the developed procedure in a real hybrid microgrid located in the facilities of the Centro de Energías Renovables (CEDER), which is part of the EU-funded TIGON project dedicated to the demonstration of innovations in hybrid microgrids for greener, more resilient and safer power grids. The measurable variables in the CEDER hybrid AC/DC microgrid have been compared with the values obtained in the power flow simulation for validation, demonstrating the accuracy and validity of the developed procedure.

The remainder of this article is organised as follows. Section \ref{s3} presents the optimal power flow formulation and the possible objective functions that can be used for optimisation. In Section \ref{s4}, KPIs used for techno-economic assessment are explained.
In Section \ref{s5}, the characteristics of TIGON CEDER demonstrator are detailed.
In Section \ref{s6}, experimental and simulation results for different operating scenarios are discussed.
Conclusions are given in Section \ref{s7}.
Finally, the microgrid model used for the AC/DC optimal power flow and the techno-economic analysis is detailed in Appendix~\ref{s2}.


\section{AC/DC Optimal Power Flow} \label{s3}

The AC/DC optimal power flow allows studying the feasibility of the microgrid operation, the self-consumption capability, the loads' supply and the power losses.

\subsection{Formulation}

The nomenclature used in the formulation of the problem is detailed at the end of the paper. 
All electrical variables are represented in phasor form.
Consider an electrical network whose topology is represented by a graph 
$\mathcal G=(\mathcal B, \mathcal L)$ where $\mathcal B=\{1, \ldots, n\}$ is
the set of buses (vertices) and 
$\mathcal L \subset \mathcal B \times \mathcal B$ the set
of lines (edges). The lines are unordered pairs of buses $(i,k)$, where
$i$ and $k$ are the pair of buses connected by the line.

A bus $k$ is adjacent to another bus $i$ if there is a line connecting them, 
i.e. if $(i,k) \in \mathcal L$. 
The set buses adjacents to bus $i$ is denoted by $\mathcal A_i$ and
is defined as follows:
\begin{align}
\mathcal A_i = \{ k \in \mathcal B \mid  (i,k) \in \mathcal L \}
\end{align}

The state of the network variables are physically related between them \citep{alexander2013fundamentos}. First, apparent power at each bus $S_i$ is expressed with Equation~\eqref{eq6}, where $I_i^*$ is the complex conjugate of $I_i$.

\begin{linenomath}
\begin{equation}
S_i = V_i \cdot I^*_i.\label{eq6}
\end{equation}
\end{linenomath}

$I_i$ is an aggregation of currents, as stated in Equation~\eqref{eq7.1}.

\begin{linenomath}
\begin{equation}
I_i = \sum_{k \in \mathcal A_i}{(I_{ik}+I_{ik0})}.\label{eq7.1}
\end{equation}
\end{linenomath}

Both $I_{ik}$ and $I_{ik0}$ can be calculated by Ohm’s Law as in Equation~\eqref{eq7} and Equation~\eqref{eq8}, respectively.

\begin{linenomath}
\begin{equation}
I_{ik} = y_{ik} \cdot (V_i-V_k),\label{eq7}
\end{equation}
\end{linenomath}

\begin{linenomath}
\begin{equation}
I_{ik0} = \frac{b_{ik} \cdot V_i}{2}.\label{eq8}
\end{equation}
\end{linenomath}

Power flow equations are obtained combining all the previous expressions. In the case of AC buses, the resulting expressions are Equation~\eqref{eq9} and Equation~\eqref{eq10} \citep{montes1995solving,samperio2023sparse}.

\begin{linenomath}
\begin{equation}
P_i = \sum_{k\in \mathcal A_i}{\left[c_{ik}\cdot(e_i^2+f_i^2-e_i\cdot e_k-f_i\cdot f_k)+s_{ik}\cdot(e_i\cdot f_k-e_k\cdot f_i) + P_{conv,ik}\right]},\label{eq9}
\end{equation}
\end{linenomath}

\begin{linenomath}
\begin{equation}
Q_i = \sum_{k\in \mathcal A_i}{\left[c_{ik}\cdot(e_i\cdot f_k-e_k\cdot f_i)+s_{ik}\cdot(-e_i^2-f_i^2+e_i\cdot e_k+f_i\cdot f_k)+b_{ik}\cdot \frac{-e_i^2-f_i^2}{2}\right]}.\label{eq10}
\end{equation}
\end{linenomath}

$P_{conv,ik}$, which is the power injected or transferred from converters, is added to the active power load flow equation.
In the case of DC buses, the expressions are Equation~\eqref{eq11} and Equation ~\eqref{eq12}.

\begin{linenomath}
\begin{equation}
P_i = \sum_{k\in \mathcal A_i}{\left[c_{ik}\cdot e_i \cdot (e_i - e_k) + P_{conv,ik}\right]},\label{eq11}
\end{equation}
\end{linenomath}

\begin{linenomath}
\begin{equation}
Q_i = 0.\label{eq12}
\end{equation}
\end{linenomath}

In any case, $P_i$ and $Q_i$ are the sum of generation and demand (Equations~\eqref{eq13} and \eqref{eq14}).

\begin{linenomath}
\begin{equation}
P_i = P_{gen,i} + P_{load,i},\label{eq13}
\end{equation}
\end{linenomath}

\begin{linenomath}
\begin{equation}
Q_i = Q_{gen,i} + Q_{load,i}.\label{eq14}
\end{equation}
\end{linenomath}

At all times $P_{gen,i}$ and $Q_{gen,i}$ are limited to the nominal power and the minimum power of the generators (Equations~\eqref{eq14.1} and \eqref{eq14.2}, respectively).

\begin{linenomath}
\begin{equation}
P_{gen,nom,i} \geq P_{gen,i},\label{eq14.1} \geq P_{gen,min,i}
\end{equation}
\end{linenomath}

\begin{linenomath}
\begin{equation}
Q_{gen,nom,i} \geq Q_{gen,i}.\label{eq14.2} \geq Q_{gen,min,i}
\end{equation}
\end{linenomath}

In case two buses are connected through a converter, the expression that describes the exchange of power between them is Equation~\eqref{eq15}.

\begin{linenomath}
\begin{equation}
P_{conv,ik} = \frac{P_{conv,ki}}{\eta_{ik}}.\label{eq15}
\end{equation}
\end{linenomath}

Maximum and minimum voltage limits of buses are respected through Inequation \ref{eq16}.

\begin{linenomath}
\begin{equation}
V_{max,i}^2 \geq e_i^2 + f_i^2 \geq V_{min,i}^2.\label{eq16}
\end{equation}
\end{linenomath}

If grid-forming mode is activated in a converter, voltage at the output bus of the converter is be set to its nominal value by Equation~\eqref{eq17}.

\begin{linenomath}
\begin{equation}
e_i^2 + f_i^2 = 1.\label{eq17}
\end{equation}
\end{linenomath}

When the converter is in grid-following mode, this restriction is not considered.

The total current that lines are able to transport is limited by Inequation \ref{eq18}.

\begin{linenomath}
\begin{equation}
I_{max,ik}^2 \geq \left[ c_{ik} \cdot (e_i - e_k) + s_{ik} \cdot (f_k - f_i) \right]^2 + \left[ c_{ik} \cdot (f_i - f_k) + s_{ik} \cdot (e_i - e_k) \right]^2.\label{eq18}
\end{equation}
\end{linenomath}

Transformers cannot exceed their nominal power when operating, as stated in Inequations \ref{eq19} and \ref{eq20}.

\begin{linenomath}
\begin{equation}
S_{n,ik}^2 \geq \left\lbrace \left[ c_{ik} \cdot (e_i - e_k) + s_{ik} \cdot (f_k - f_i) \right]^2 + \left[ c_{ik} \cdot (f_i - f_k) + s_{ik} \cdot (e_i - e_k) \right]^2\right\rbrace \cdot (e_i^2 + f_i^2),\label{eq19}
\end{equation}
\end{linenomath}

\begin{linenomath}
\begin{equation}
S_{n,ik}^2 \geq \left\lbrace \left[ c_{ik} \cdot (e_i - e_k) + s_{ik} \cdot (f_k - f_i) \right]^2 + \left[ c_{ik} \cdot (f_i - f_k) + s_{ik} \cdot (e_i - e_k) \right]^2\right\rbrace \cdot (e_k^2 + f_k^2).\label{eq20}
\end{equation}
\end{linenomath}

In the case of converters, this limitation is expressed by Inequation \ref{eq21}.

\begin{linenomath}
\begin{equation}
S_{n,ik} \geq P_{conv,ik}.\label{eq21}
\end{equation}
\end{linenomath}

\subsection{Optimisation}

Four different objective functions $h$ can be chosen for being minimised in the AC/DC optimal power flow, depending on the objective to be achieved:

\begin{enumerate}
\item \textbf{H1: Total active power generated.} Focuses on reducing energy losses (Equation~\eqref{eq23}).

\begin{linenomath}
\begin{equation}
H1 = \sum_{i\in\mathcal B}{P_{gen,i}}.\label{eq23}
\end{equation}
\end{linenomath}

\item \textbf{H2: Buses voltage deviation from their nominal value.} Focuses on achieving grid stability (Equation~\eqref{eq24}).

\begin{linenomath}
\begin{equation}
H2 = \sum_{i\in\mathcal B}{(e_i^2 + f_i^2 - 1)^2}.\label{eq24}
\end{equation}
\end{linenomath}

\item \textbf{H3: Total amount of operational costs associated to each generator.} Focuses in achieving economic savings (Equation~\eqref{eq25}).

\begin{linenomath}
\begin{equation}
H3 = \sum_{i\in\mathcal B}{OC_{i} \cdot P_{gen,i}}.\label{eq25}
\end{equation}
\end{linenomath}

\item \textbf{H4: Active power microgrid generation.} Focuses on making the highest possible use of the available generation resources in the microgrid (Equation~\eqref{eq22}). The objective function is negative in order to calculate a maximization.

\begin{linenomath}
\begin{equation}
H4 = -\sum_{i\in\mathcal B}{P_{gen,i}}.\label{eq22}
\end{equation}
\end{linenomath}

\end{enumerate}

The variables that are optimised when performing the power flow are $e_i$, $f_i$, $P_{gen,i}$, $Q_{gen,i}$, and $P_{conv,ik}$.


\section{Techno-economic Assessment} \label{s4}

The techno-economic assessment presented in this paper consists in the calculation of eight KPIs. They allow evaluating a microgrid in terms of costs, energy generation, storage capabilities and financial feasibility. Morover, it provides critical insights for making informed decisions and maximizing the overall performance of the microgrid. The KPIs, presented below, are divided into two categories: technical and economic.

\subsection{Technical KPIs}

\begin{itemize}

\item \textbf{KPI1: Electrical energy generated.}
Amount of electrical energy generated by year.

\begin{linenomath}
\begin{equation}
KPI1 \ (kWh) = \sum_{i\in\mathcal B}{P_{gen,nom,i} \cdot CF_i \cdot 3600 h}.\label{eq26}
\end{equation}
\end{linenomath}

\item \textbf{KPI2: CO\textsubscript{2} emissions.}
Total CO\textsubscript{2} emitted by all energy carriers associated with the primary energy use in the microgrid by year.

\begin{linenomath}
\begin{equation}
KPI2 \ (kgCO_2) = \sum_{i\in\mathcal B}{P_{gen,nom,i} \cdot CF_i \cdot 3600 h \cdot GHG_{i}}.\label{eq27}
\end{equation}
\end{linenomath}

\item \textbf{KPI3: Self-consumption percentage.}
Amount of energy obtained from the generators of the microgrid in relation to the energy used by the loads.

\begin{linenomath}
\begin{equation}
KPI3 \ (\%) = \frac{\sum_{i\in\mathcal B}{P_{gen,nom,i} \cdot CF_i}}{\sum_{i\in\mathcal B}{P_{load,i}}}.\label{eq28}
\end{equation}
\end{linenomath}

\item \textbf{KPI4: Storage flexibility.}
Total flexible power available in the microgrid thanks to the storage systems.

\begin{linenomath}
\begin{equation}
KPI4 \ (kW) = \sum_{i\in\mathcal B}{P_{stor,i}}.\label{eq29}
\end{equation}
\end{linenomath}

\end{itemize}

\subsection{Economic KPIs}

\begin{itemize}

\item \textbf{KPI5: Total life cycle income.}
Total income earned by the microgrid along its useful life.

\begin{linenomath}
\begin{equation}
KPI5 \ (currency) = {\sum_{n=1}^{UL}{\frac{\sum_{i\in\mathcal B}{(P_{gen,nom,i} \cdot CF_i \cdot EP)} + FI}{(1 + r)^n}}}.\label{eq30}
\end{equation}
\end{linenomath}

\item \textbf{KPI6: Total life cycle cost.}
Total cost incurred by the microgrid along its useful life.

\begin{linenomath}
\begin{equation}
KPI6 \ (currency) = IC + \sum_{i\in\mathcal B} {(IC_i + OC_i)} + \sum_{n=1}^{UL} {\frac{OMC + \sum_{i\in\mathcal B} {MC_i}} {(1 + r)^n}} - \frac{RV - \sum_{i\in\mathcal B} {RV_i}}{(1 + r)^{UL}}.\label{eq31}
\end{equation}
\end{linenomath}

\item \textbf{KPI7: Payback.}
Period of time required to recover the capital investment of the microgrid \citep{kiran2022principles}.

\begin{multline}
KPI7 \ (years) = N \ when ({\sum_{n=1}^{N}{\frac{(\sum_{i\in\mathcal B}{P_{gen,nom,i} \cdot CF_i \cdot EP)}+FI}{(1 + r)^n}}} = \\
= IC + \sum_{i\in\mathcal B} {(IC_i + OC_i)} + \sum_{n=1}^{UL} {\frac{OMC + \sum_{i\in\mathcal B} {MC_i}} {(1 + r)^n}} - \frac{RV - \sum_{i\in\mathcal B} {RV_i}}{(1 + r)^{UL}}).\label{eq32}
\end{multline}

\item \textbf{KPI8: Levelized cost of energy (LCOE).}
Price at which the generated electricity should be sold to break even at the end of the microgrid's useful life \citep{papapetrou2017assessment, abadie2019levelized}.

\begin{linenomath}
\begin{equation}
KPI8 \ (currency/kWh) = \frac{IC + \sum_{i\in\mathcal B} {(IC_i + OC_i)} + \sum_{n=1}^{UL} {\frac{OMC + \sum_{i\in\mathcal B} {MC_i}} {(1 + r)^n}} - \frac{RV - \sum_{i\in\mathcal B} {RV_i}}{(1 + r)^{UL}}}{\sum_{n=1}^{UL}{\frac{\sum_{i\in\mathcal B}{P_{gen,nom,i} \cdot CF_i}}{(1 + r)^n}}}.\label{eq33}
\end{equation}
\end{linenomath}

\end{itemize}


\section{TIGON CEDER Demonstrator} \label{s5}

\begin{figure}[H]
\centering
\includegraphics[width=.45\textwidth]{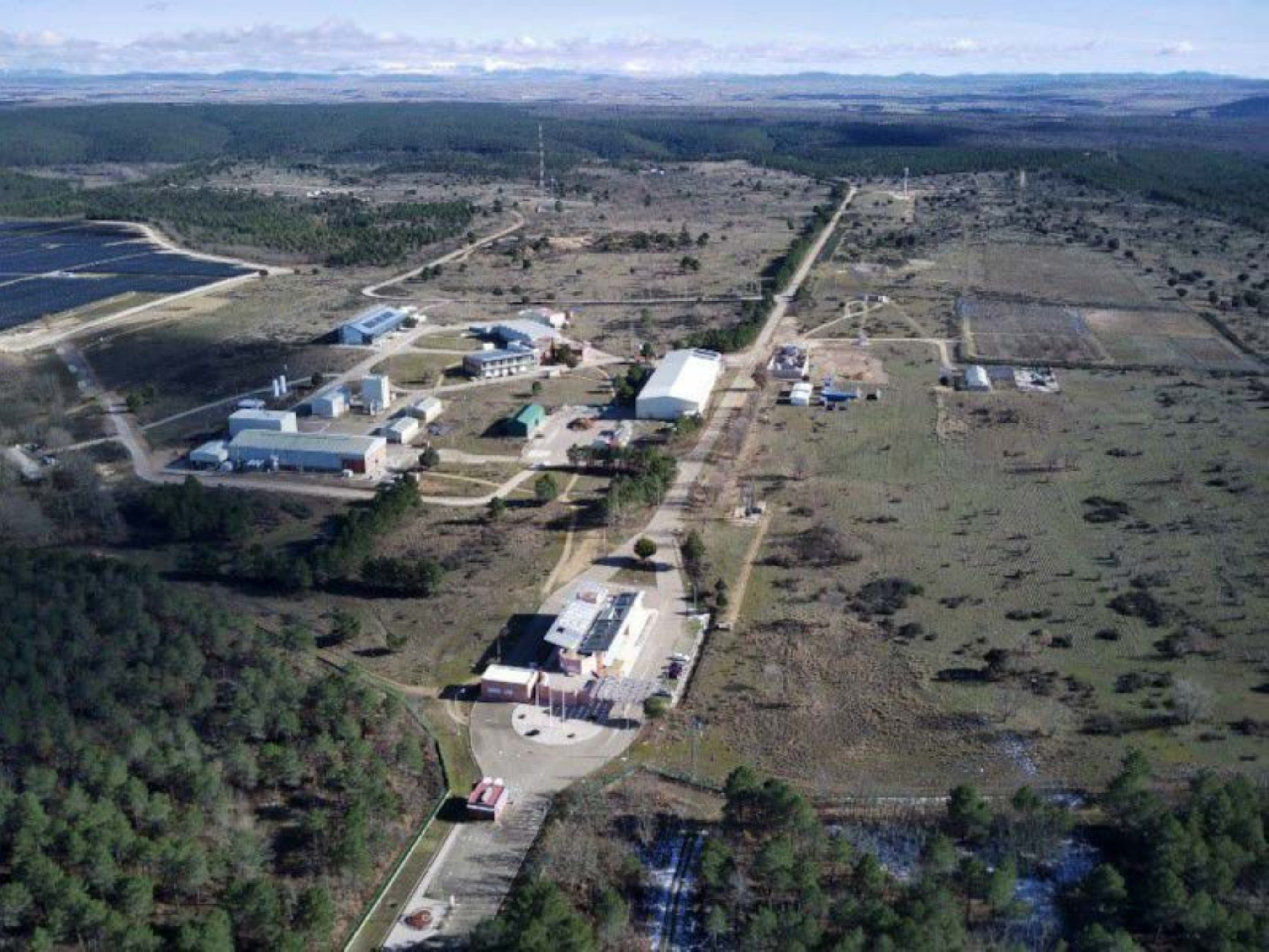}
\includegraphics[width=.45\textwidth]{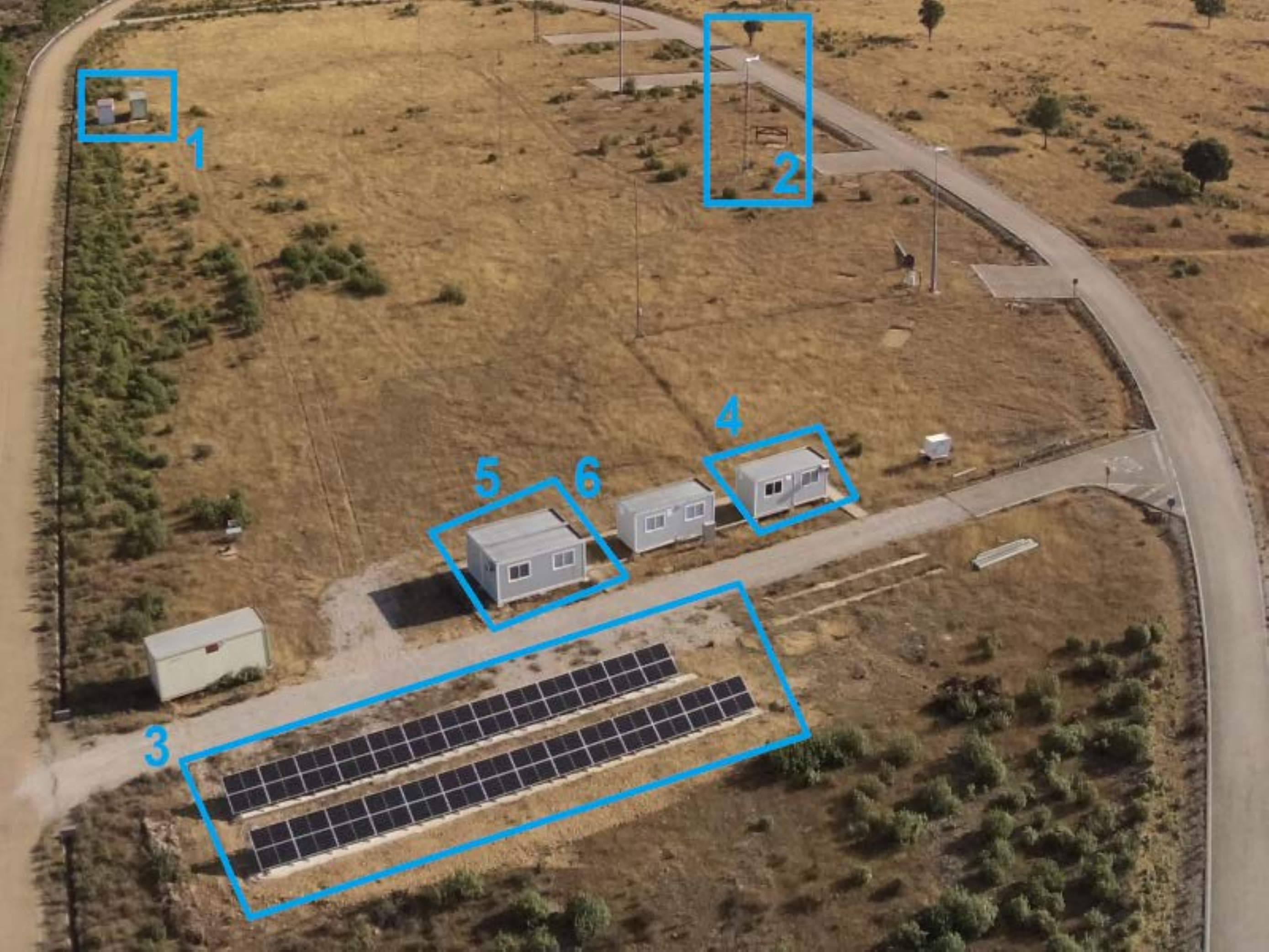}
\caption{CEDER facilities (left) and  TIGON demostrator (right).\label{fig:TIGON}}
\end{figure}

CEDER is the acronym for Centre for the Development of Renewable Energies \footnote{http://www.ceder.es/}. Is located in Lubia (Soria, Spain), and is a national centre for energy research that belongs to the Centre for Energy, Environmental and Technological Research (CIEMAT) \footnote{https://www.ciemat.es/portal.do?IDM=6\&NM=1}, a Public Research Organisation dependent on the Ministry of Science and Innovation.
CEDER covers an area of 640 ha (13,000 m\textsuperscript{2} built) and has an smart microgrid (electrical and thermal), operated and managed in real time (see Figure~\ref{fig:TIGON}, left).

The Spanish demonstrator of TIGON is installed at CEDER (see Figure~\ref{fig:TIGON}, right) and consists of the following elements:

\begin{enumerate}
\item Transformer station (15 kV\textsubscript{AC} - 400 V\textsubscript{AC}).
\item Small wind turbine: A three-bladed horizontal axis wind turbine with a nominal power of 4.2kW (Ryse E5).
\item Photovoltaic system (PV): Three strings with 18 modules Ureco of 410 W. Total 22.14 kW.
\item NMC batteries: Three modules of 80 cells (50 Ah and 3.6 V each cell).
\item Programmable AC loads: Three programmable AC loads of 2.9 kW each one.
\item DC loads: Three adjustable DC loads of 4 kW each one.
\end{enumerate}

\begin{figure}[H]
\centering
\includegraphics[width=11.5cm]{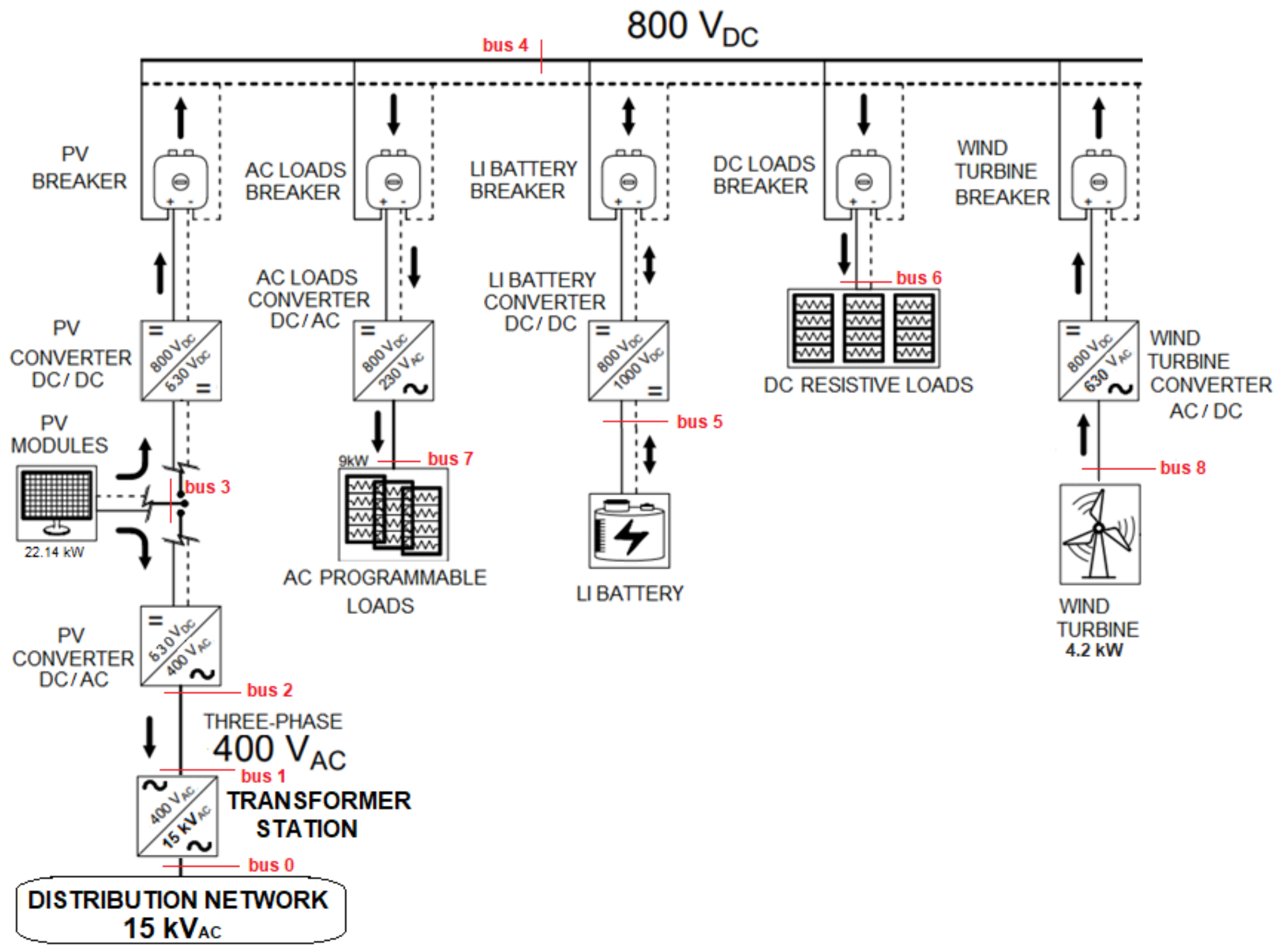}
\caption{TIGON layout at CEDER.\label{cederer}}
\end{figure}

A schematic of the CEDER demonstrator is shown in Figure~\ref{cederer}. It is constituted by AC loads, wind turbines, transformers and DC sections in the network. The characteristics of the elements with which the microgrid have been tested are collected in Tables~\ref{tab:Edata}-\ref{tab7}.

$V_{max,i}$ and $V_{min,i}$ have been fixed at five per cent above and below the nominal value, respectively. $GHG_i$ of generators 0 and 1 have been obtained form \citep{baldwin2006carbon}. Generators 2 and 3 models the reactive power management of converters 0 and 3, respectively. The CEDER microgrid has one connection to an external grid located on bus 0. The frequency is 50~Hz ($\omega = 100\pi$~rad/s).

$IC$ and $RV$ have been calculated as the sum of the investment cost and the residual value of the microgrid equipment and labour listed in Table~\ref{tab:equip}.


\section{Results And Discussion} \label{s6}

\subsection{AC/DC Optimal Power Flow}

The AC/DC optimal power flow simulation has been implemented in Python due to its wide range of available open source optimisation libraries. In this paper, the optimisation has been performed using CasADi \citep{Andersson2019}, which is a software library equipped with specific tools focused on modelling, optimisation and control of nonlinear dynamic systems. CasADi is widely used to define both mathematical models and constraints involved, and allows utilising different solvers in order to optimise the problem. In this study case, the solver Ipopt \citep{wachter2006implementation} has been used. Ipopt applies a sequential quadratic programming (SQP) to solve constrained nonlinear optimization problems, which is the case of the AC/DC optimal power flow. The calculation time is not significant, it is only a few seconds.

Four scenarios have been tested, one per each objective function. In all scenarios, TIGON CEDER storage has been considered to act as a load that consumes electricity at half of its nominal power ($P_{stor}$/2), and the external grid has been limited to power consumption (if required). The results obtained at each scenario are included in Table \ref{tab:simu}.

In all scenarios, all the power demanded by loads and storages are delivered. In none of them exists demand of reactive power, so all the reactive power in the microgrid is generated in the AC section at bus 2 because of the transformer and AC line reactance.

In scenarios H1 and H3 the exact active power is generated to supply the demand and compensate losses. The difference between them is that the wind turbine reduce its active power generation in scenario H3 because this technology has highest operational costs than the PV, whereas more power is saved in scenario H1.

In scenario H2, $V_i$ at all buses are the nominal value or very close to it. To achieve it, the generation of active power and the consumption of the external grid is precisely adjusted.

In scenario H4, the PV and the wind turbine generate as much active power as possible, resulting in supplying power to the external grid.

All the scenarios have been recreated in the real environment of the TIGON CEDER microgrid. The two quantities that have been accessed for measurement are active power and voltage. The values obtained in the measurements taken for each scenario are shown in Table~\ref{tab:meas}.

The measures obtained are very similar to the calculated values in the simulations. The largest differences are between the expected and actually measured PV voltages, which are at most 3\% error. This occurs because the PV voltage (bus 3) varies considerably as the delivered active power changes. Little differences come from the accuracy of the monitoring devices and the difficulty of precisely obtain at the same time the proposed values of active power generation in both the wind turbine and the PV, due to their inherent variability depending on the weather conditions and technical restrictions.

The results confirm that the AC/DC optimal power flow works correctly and accurately enough to assess the operation of the microgrid.

\subsection{Techno-economic Assessment}

The techno-economic assessment calculations have been implemented in Python. Again, the calculation time is very short, only a few seconds.
Four scenarios in which the TIGON CEDER microgrid could operate have been studied: baseline, no battery, battery flexibility and virtual battery (VB) flexibility.

In the baseline scenario, the microgrid elements have been considered with the same characteristics presented in Section \ref{s5}. Nevertheless, in this situation, which is the real one, the battery is being used only for performing test and doing research experiments. For this reason, we have considered three more scenarios in which the microgrid could be more profitable.
In the no-battery scenario, the battery and its converter have been removed from the microgrid, and so their investment costs and residual values.
In the battery flexibility scenario, the battery has not been eliminated, but it has been deemed for being used in the Spanish upward tertiary regulation market. Batteries are loads with inherent electrical flexibility, as they can be charged and discharged at the most convenient time, keeping the daily balance of generated and consumed electricity unchanged. The upward tertiary regulation market has been chosen because it is the balance market with the highest average price in Spain in 2022: 224.17 \euro{}/MWh \footnote{https://www.sistemaelectrico-ree.es/informe-del-sistema-electrico/mercados/servicios-ajuste/energias-precios-balance}.
The power of the battery is very small to participate in the Spanish tertiary regulation market on its own, as it is necessary to make bids of at least 1 MW \citep{REEguia}. Therefore, it needs to be part of an aggregation that participates in the market as a unitary market agent, following the methodology explained in \citep{martin2023aggregated}.
Lastly, in the VB flexibility scenario the battery is replaced by a VB consisting of an aggregation of thermostatically controlled loads (TCLs) with the same nominal power $P_{stor,i}$. These TCLs could be heat pumps or water heaters, for instance. The aggregation of TCLs in VB is achieved using the method proposed in \citep{martin2021flexibility}.
For both flexibility scenarios, it is assumed a storage participation in the market for one hour per day operating at its nominal power $P_{stor,i}$ and that the aggregation shares its benefits proportionally between its loads. This results on an additional flexibility income $FI$ of 2,045.55 \euro{} per year.

The KPIs values obtained for each scenario are presented in Table~\ref{tab:kpi}.

KPI 2 (CO\textsubscript{2}) proves that carbon dioxide emissions are very low. This is caused by the usage of renewable generation technologies, specifically PV and wind turbine.
KPI 3 (self-consumption percentage) shows that the microgrid produces more energy than the required by the loads per year, thus being able to compensate energy losses.

KPI 7 (payback) shows that in the baseline scenario the microgrid is not profitable. It make sense, as the TIGON CEDER microgrid is a small-scale network where the elements are high-cost prototypes. The battery, which is the storage element of the grid, is key for increasing or decreasing costs and incomes. The battery flexibility scenario demonstrates that using the battery for making bids in the Spanish electricity balance markets allows increasing the incomes and making the microgrid profitable. Nevertheless, the flexibility remuneration is not high enough to make the microgrid payback lower than in the case that the battery is removed, i.e. the no battery scenario. In the no battery scenario, costs are highly reduced, as well as KPI 8 (LCOE). This is on the expense of reducing KPI 4 (storage flexibility) to zero.
The most profitable scenario is VB flexibility, as it combines the additional incomes obtained by the participation in the tertiary regulation market with the cost savings caused by the disappearance of the physical battery. KPI 7 (payback) is reduced to 17 years and KPI 8 (LCOE) to 113.92 \euro{}/kWh, whereas KPI 4 (storage flexibility) remains in 25 kW.


\section{Conclusions} \label{s7}

In this paper, an AC/DC optimal power flow and a techno-economic assessment has been presented with the aim of helping users evaluating the operation and the viability of their hybrid AC/DC microgrids. In addition, both calculations can be used for proposing improvements and new investments. The AC/DC optimal power flow is an useful technique for checking the correct operation of hybrid microgrids under specific instantaneous conditions, while the techno-economic assessment allows to verify its performance in the long term.

Both methodologies have been tested on the TIGON CEDER microgrid, which has been described in this paper. The AC/DC optimal power flow has been validated using real measurements and considering different objective function scenarios.
Between the four techno-economic scenarios, the VB flexibility one is the most profitable and the one with the fastest return on investment. Based on the results of flexibilities scenarios, we encourage the actors involved in the Spanish electricity system to increase the remuneration of flexibility, so as to increase the penetration of renewable energy sources and advance in the transition to a more efficient AC/DC hybrid power system with low greenhouse gas emissions. 

As future work, we will apply the AC/DC optimal power flow and the techno-economic assessment to other microgrids, especially those primarily aimed at supplying electricity to residential consumers and industries. We also hope that the two developed techniques and the TIGON CEDER microgrid example can be used by other companies and research institutions in their developments.


\nomenclature{$\mathcal G$}{Microgrid topology network graph}
\nomenclature{$\mathcal B$}{Set of buses}
\nomenclature{$\mathcal L$}{Set of lines}
\nomenclature{$j$}{Imaginary unit}
\nomenclature{$i$, $k$}{Subindices denoting buses}
\nomenclature{$n$}{Subindex denoting years}
\nomenclature{$z_{ik}$}{Line impedance ($\Omega$)}
\nomenclature{$r_{ik}$}{Line resistance ($\Omega$)}
\nomenclature{$x_{ik}$}{Line reactance ($\Omega$)}
\nomenclature{$C_{ik}$}{Line capacitance (F)}
\nomenclature{$y_{ik}$}{Line admittance ($\Omega^{-1}$)}
\nomenclature{$c_{ik}$}{Line conductance ($\Omega^{-1}$)}
\nomenclature{$s_{ik}$}{Line susceptance ($\Omega^{-1}$)}
\nomenclature{$b_{ik}$}{Line susceptance to ground ($\Omega^{-1}$)}
\nomenclature{$I_{max,ik}$}{Line current limit (A)}
\nomenclature{$\omega$}{Microgrid angular frequency (rad/s)}
\nomenclature{$S_i$}{Bus apparent power (VA)}
\nomenclature{$V_i$}{Bus voltage (V)}
\nomenclature{$V_{n,i}$}{Bus nominal voltage (V)}
\nomenclature{$V_{max,i}$}{Bus maximum voltage (V)}
\nomenclature{$V_{min,i}$}{Bus minimum voltage (V)}
\nomenclature{$e_i$}{Bus real part of voltage (V)}
\nomenclature{$f_i$}{Bus imaginary part of voltage (V)}
\nomenclature{$\delta_i$}{Bus voltage phase (rad)}
\nomenclature{$I_i$}{Bus injected current (A)}
\nomenclature{$I_{ik}$}{Current flowing from i to k or vice versa (A)}
\nomenclature{$I_{max,ik}$}{Maximum current flowing from i to k or vice versa (A)}
\nomenclature{$I_{ik0}$}{Current flowing to ground between i and k (A)}
\nomenclature{$P_i$}{Bus total active power (W)}
\nomenclature{$Q_i$}{Bus total reactive power (VAr)}
\nomenclature{$V_{LN,ik}$}{Transformer nominal voltage (V)}
\nomenclature{$V_{ccL,ik}$}{Transformer zero sequence relative short-circuit voltage percentage (\%)}
\nomenclature{$V_{RccL,ik}$}{Transformer resistive part of the zero sequence relative short-circuit voltage percentage (\%)}
\nomenclature{$S_{n,ik}$}{Transformer or converter nominal power (VA)}
\nomenclature{$P_{conv,ik}$}{Converter injected or transferred active power (W)}
\nomenclature{$\eta_{ik}$}{Converter performance}
\nomenclature{$P_{gen,i}$}{Generator active power (W)}
\nomenclature{$Q_{gen,i}$}{Generator reactive power (W)}
\nomenclature{$P_{gen,nom,i}$}{Generator nominal active power (W)}
\nomenclature{$Q_{gen,nom,i}$}{Generator nominal active power (VAr)}
\nomenclature{$P_{gen,min,i}$}{Generator minimum active power (W)}
\nomenclature{$Q_{gen,min,i}$}{Generator minimum active power (VAr)}
\nomenclature{$IC_{i}$}{Generator investment cost (currency)}
\nomenclature{$RV_{i}$}{Generator residual value (currency)}
\nomenclature{$MC_{i}$}{Generator maintenance cost (currency/year)}
\nomenclature{$OC_{i}$}{Generator operational cost (currency/kWh)}
\nomenclature{$CF_{i}$}{Generator capacity factor (\%)}
\nomenclature{$GHG_{i}$}{Generator CO\textsubscript{2} equivalent (kgCO\textsubscript{2}/kWh)}
\nomenclature{$P_{load,i}$}{Load active power (W)}
\nomenclature{$Q_{load,i}$}{Load reactive power (W)}
\nomenclature{$P_{stor,i}$}{Storage nominal power (W)}
\nomenclature{$IC$}{Microgrid investment cost (currency)}
\nomenclature{$RV$}{Microgrid residual value (currency)}
\nomenclature{$OMC$}{Microgrid operation and maintenance cost (currency/year)}
\nomenclature{$UL$}{Microgrid useful life (years)}
\nomenclature{$r$}{Discount rate (\%)}
\nomenclature{$EP$}{Electricity price (currency/kWh)}
\nomenclature{$FI$}{Flexibility income (currency)}
\nomenclature{$S_{base}$}{Base power (VA)}
\nomenclature{$V_{base}$}{Base voltage (V)}
\nomenclature{$z_{base}$}{Base impedance ($\Omega$)}

\printnomenclature

\section*{Conflict of Interest Statement}

The authors declare that the research was conducted in the absence of any commercial or financial relationships that could be construed as a potential conflict of interest.

\section*{Author Contributions}

AM-C: Conceptualization, Data curation, Formal analysis, Investigation, Methodology, Software, Validation, Visualization, Writing – original draft, Writing – review \& editing. AH-S: Conceptualization, Formal analysis, Investigation, Methodology, Project administration, Software, Writing – review \& editing. OI-M: Conceptualization, Data curation, Formal analysis, Investigation, Project administration, Resources, Validation, Visualization, Writing – review \& editing. PP-C: Investigation, Resources, Writing – review \& editing. AH-J: Investigation, Resources, Writing – review \& editing. FF-E: Conceptualization, Methodology, Supervision, Writing – review \& editing. EB: Conceptualization, Methodology, Supervision, Writing – review \& editing.

\section*{Funding}
This research has received funding from the European Union’s Horizon 2020 TIGON project under grant agreement No 957769.

\appendix
\section[]{Appendix: Microgrid Model} \label{s2}

\begin{figure}[H]
\centering
\includegraphics[width=9.5cm]{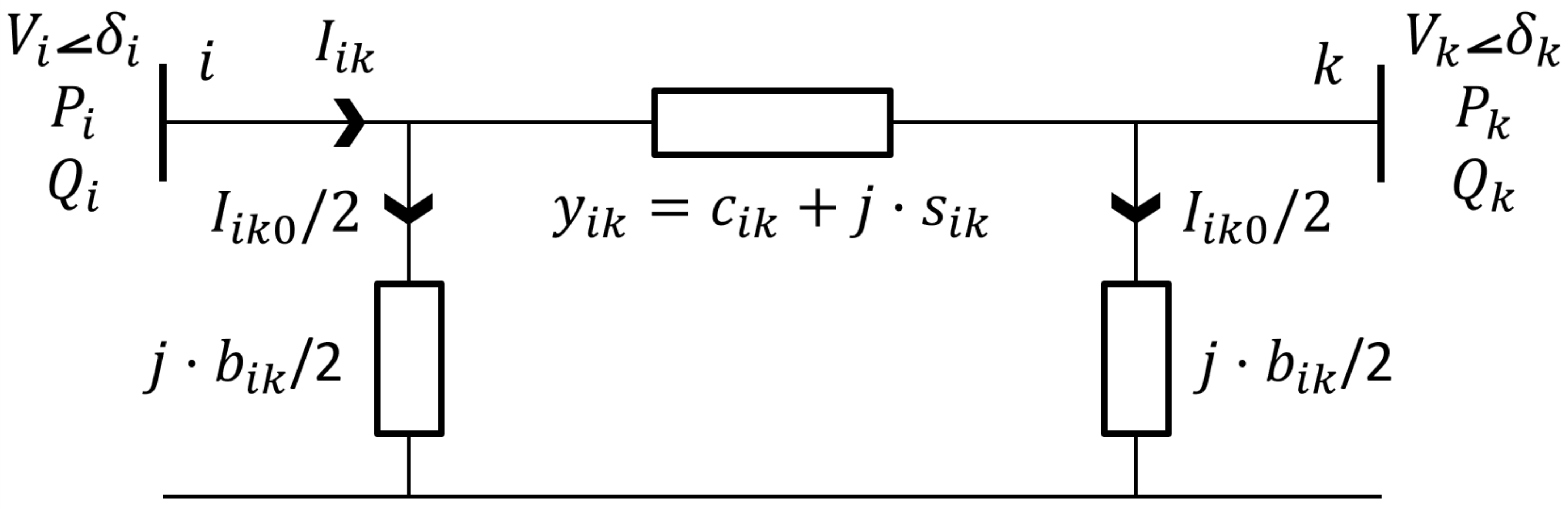}
\caption{$\pi$ model of lines.\label{pimodel}}
\end{figure}

The $\pi$ model representation of grid lines (Figure \ref{pimodel} \citep{cui2017static,alexander2013fundamentos}) is used for performing the AC/DC optimal power flow. In the case of DC lines, $s_{ik}$ and $b_{ik}$ are not considered, as the reactive part of impedances does not affect electric charges when they move always in the same direction over time.

$y_{ik}$ is the inverse of the line impedance $z_{ik}$, which is expressed by Equation~\eqref{eq1}.

\begin{linenomath}
\begin{equation}
z_{ik} = r_{ik} + j \cdot x_{ik}.\label{eq1}
\end{equation}
\end{linenomath}

Both $r_{ik}$ and $x_{ik}$ are directly proportional to the line length, and, in the case of $x_{ik}$, also the frequency. In the case of DC lines, $x_{ik}$ is not considered because it is a reactive parameter. For AC cases, $b_{ik}$ is obtained using Equation~\eqref{eq2}.

\begin{linenomath}
\begin{equation}
b_{ik} = \omega \cdot C_{ik}.\label{eq2}
\end{equation}
\end{linenomath}

$C_{ik}$ is directly proportional to the line length. Each line has a current limit $I_{max,ik}$.

Voltage at each bus can be expressed by its real and imaginary part, as in Equation~\eqref{ghty}

\begin{linenomath}
\begin{equation}
V_{i} = e_i + j \cdot f_i.\label{ghty}
\end{equation}
\end{linenomath}

The bus voltage phase $\delta_i$ can be calculated by Equation~\eqref{Roronoa}.

\begin{linenomath}
\begin{equation}
\delta_i = \arctan{\frac{f_i}{e_i}}.\label{Roronoa}
\end{equation}
\end{linenomath}

Each bus of the grid can have a different reference voltage. In order to ease load flow calculations in grids where voltage and current transformations occur through converters or transformers, a base power $S_{base}$ and a base voltage $V_{base}$ are set. Base impedance $z_{base}$ is calculated as in Equation~\eqref{eq3}.

\begin{linenomath}
\begin{equation}
z_{base} = \frac{V_{base}^2}{S_{base}}.\label{eq3}
\end{equation}
\end{linenomath}

All magnitudes in the grid model are divided by their base values, so they can be expressed per unit (p.u.). Maximum and minimum voltages at each bus are specified ($V_{max}$ and $V_{min}$, respectively).

Transformers in the grid are modelled in the same way as lines, but without taking $b_{ik}$ into account because it is considered to have a minimal impact on the total transformer impedance. The magnitude and the resistive part of their $z_{ik}$ are calculated with Equations~\eqref{eq4} and \eqref{eq5}.

\begin{linenomath}
\begin{equation}
|z_{ik}| = \frac{V_{ccL,ik} \cdot V_{LN,ik}^2}{100 \cdot S_{n,ik}},\label{eq4}
\end{equation}
\end{linenomath}

\begin{linenomath}
\begin{equation}
r_{ik} = \frac{V_{RccL,ik} \cdot V_{LN,ik}^2}{100 \cdot S_{n,ik}}.\label{eq5}
\end{equation}
\end{linenomath}

Besides transformers, converters are the other grid elements that changes voltage levels between buses. There exists AC/DC, DC/AC, AC/AC and DC/DC converters \citep{mohammed2021state}. Their characteristics needed to perform load flow calculations are nominal power $S_{n,ik}$, performance $\eta_{ik}$ and control mode (grid-following or grid-forming). The reactive power at the output of DC/AC and AC/AC converters is included in the grid model of this paper as a reactive power generator in order to recreate the reactive power management operation of these converters.

All generators existing in the microgrid are characterized by their nominal active power $P_{gen,nom,i}$, minimum active power $P_{gen,min,i}$, nominal reactive power $Q_{gen,nom,i}$ and minimum reactive power $Q_{gen,min,i}$.
Regarding loads, they are defined by two parameters: demanded active power $P_{load,i}$ and demanded reactive power $Q_{load,i}$. Supplying $P_{load,i}$ and $Q_{load,i}$ is considered mandatory.
In the case of storages, they can act as generators of loads in the power flow. Their defining parameter is $P_{stor,i}$.

If present, external grids are considered to be capable of providing and consuming active power and reactive power with no limits, but not at the same time.

For the techno-economic assessment, additional information about generators is considered. This includes their investment cost $IC_{i}$, residual value $RV_{i}$, maintenance cost $MC_{i}$, operational cost $OC_{i}$, capacity factor $CF_{i}$ and CO\textsubscript{2} equivalent $GHG_{i}$. $CF_{i}$ is the percentage of power produced by the generator with respect to the power that could have been produced if the generator had operated at maximum rate \citep{quezada2006assessment}.

Regarding the rest of the microgrid, investment cost $IC$, residual value $RV$, operation and maintenance cost $OMC$ and useful life $UL$ are used. In addition, the price at which the microgrid sells electricity $EP$,and the discount rate $r$ are included in the model. Flexibility income $FI$ is considered in the flexibility scenario of the techno-economic analysis.

\bibliographystyle{Frontiers-Harvard} 
\bibliography{Bibliografia_TIGON}


\newpage

\section*{Figure and table captions}


\begin{table}[H]
\caption{TIGON CEDER electrical data.\label{tab:Edata}}
\begin{adjustbox}{max width=\textwidth}
\begin{tabular}{cccccc}
\toprule
\textbf{Bus} & \textbf{$V_{n,i}$ (kV)} & \textbf{Type} \\
\midrule
0 & 15.000 & AC \\
1 & 0.400 & AC \\
2 & 0.400 & AC \\
3 & 0.630 & DC \\
4 & 0.800 & DC \\
5 & 0.860 & DC \\
6 & 0.800 & DC \\
7 & 0.230 & AC \\
8 & 0.630 & AC \\
\toprule
\textbf{Line} & \textbf{$(i,k)$} &
\textbf{Length (km)} & \textbf{$r_{ik}$ ($\Omega$/km)} & 
\textbf{$x_{ik}$ ($\Omega$/km)} \\
\midrule
0 & (2,1) & 0.15 & 0.5 & 0.35 \\
1 & (4,6) & 0.20 & 0.5 & -- \\
\toprule
\textbf{Line} & \textbf{$(i,k)$} &
\textbf{$c_{ik}$ (nF/km)} & \textbf{$I_{max,ik}$ (kA)} & \textbf{Type} \\
\midrule
0 & (2,1) & 100 & 100 & AC \\
1 & (4,6) & -- & 100 & DC \\
\toprule
\textbf{Transf.}	& \textbf{$(i,k)$} & \textbf{$S_{n,ik}$ (kVA)} & \textbf{$V_{ccL,ik}$ (\%)} & \textbf{$V_{RccL,ik}$ (\%)} \\
\midrule
0 & (1,0) & 250 & 1.5 & 1.0 \\
\toprule
\textbf{Converter} & \textbf{$(i,k)$} & \textbf{$S_{n,ik}$ (kVA)} & \textbf{$\eta_{ik}$} & \textbf{Control} \\
\midrule
0 & (3,2) & 20 & 0.99 & grid-following \\
1 & (3,4) & 20 & 0.99 & grid-forming \\
2 & (4,5) & 30 & 0.86 & grid-forming \\
3 & (4,7) & 12 & 0.99 & grid-forming \\
4 & (8,4) & 5 & 0.99 & grid-following \\
\toprule
\textbf{Generator} & \textbf{Bus $i$}& \textbf{$P_{gen,max,i}$ (kW)} & \textbf{$P_{gen,min,i}$ (kW)} & \textbf{$Q_{gen,max,i}$ (kW)} & \textbf{$Q_{gen,min,i}$ (kW)} \\
\midrule
0 & 3 & 22.14 & 0.5 & 0 & 0 \\
1 & 8 & 4.20 & 0.0 & 10 & -10 \\
2 & 2 & 0.00 & 0.0 & 10 & -10 \\
3 & 7 & 0.00 & 0.0 & 10 & -10 \\
\toprule
\textbf{Load} & \textbf{Bus $i$} & \textbf{$P_{load,i}$ (kW)} & \textbf{$Q_{load,i}$ (kW)} \\
\midrule
0 & 6 & 5 & 0 \\
1 & 7 & 4 & 0 \\
\toprule
\textbf{Storage} & \textbf{Bus $i$} & \textbf{$P_{stor,i}$ (kW)} \\
\midrule
0 & 5 & 25 \\
\bottomrule
\end{tabular}
\end{adjustbox}
\end{table}

\begin{table}[H]
\caption{Techno-economic information of TIGON CEDER generators.\label{tab6}}
\begin{tabular}{ccccc}
\toprule
\textbf{Generator} & \textbf{Bus $i$} & \textbf{$IC_{i}$ (\euro{})} & \textbf{$RV_{i}$ (\euro{})} & \textbf{$MC_{i}$ (\euro{}/year)} \\
\midrule
0 & 3 & 23,000 & 100 & 200 \\
1 & 8 & 10,000 & 150 & 600 \\
\toprule
\textbf{Generator} & \textbf{Bus $i$} & \textbf{$OC_{i}$ (\euro{}/kWh)} & \textbf{$CF_{i}$ (\%)} & \textbf{$GHG_i$ (kgCO\textsubscript{2}/kWh)}\\
\midrule
0 & 3 & 0.003 & 33.5 & 0.03500 \\
1 & 8 & 0.008 & 42.0 & 0.00464 \\
\bottomrule
\end{tabular}
\end{table}

\begin{table}[H]
\caption{Techno-economic information of TIGON CEDER microgrid.\label{tab7}}
\begin{tabular}{cccccc}
\toprule
\textbf{$IC$ (\euro{})}	& \textbf{$RV$ (\euro{})} & \textbf{$OMC$ (\euro{}/year)} & \textbf{$r$ (\%)} & \textbf{$UL$ (years)} & \textbf{$EP$ (\euro{}/kWh)} \\
\midrule
195,500 & 28,900 & 1,400 & 1 & 25 & 0.145 \\
\bottomrule
\end{tabular}
\end{table}

\begin{table}[H]  
\caption{Investment costs and residual values of TIGON CEDER equipment and labour.\label{tab:equip}}
\begin{tabular}{ccc}
\toprule
\textbf{Equipment}	& \textbf{Investment cost (\euro{})} & \textbf{Residual value (\euro{})} \\
\midrule
DC PV converter & 23,000 & 2,500 \\
AC PV converter & 22,000 & 2,000 \\
Wind turbine converter & 10,000 & 500 \\
Battery & 34,000 & 400 \\
Battery converter & 25,000 & 3,000 \\
AC loads converter & 31,000 & 2,500 \\
Wiring & 15,500 & 4,000 \\
Cabins & 18,000 & 18,000 \\
Labour & 17,000 & 0 \\

\bottomrule
\end{tabular}
\end{table}

\begin{table}[H]
\caption{Simulation results for different scenarios.\label{tab:simu}}
\begin{tabular}{cccccc}
\toprule
\textbf{Bus $i$} & \textbf{Scenario} & \textbf{$P_i$ (kW)} & \textbf{$Q_i$ (kVA)} & \textbf{$V_i$ (kV)} & \textbf{$\delta_i$ (rad)} \\
\midrule
0 & H1 & 0.00000 & 0.00000 & 14.82807 & -2.42520 \\
1 & H1 & 0.00000 & 0.00000 & 0.39542 & -2.42520 \\
2 & H1 & 0.00000 & -0.00074 & 0.39542 & -2.42519 \\
3 & H1 & 19.75450 & 0.00000 & 0.62046 & 0.00000 \\
4 & H1 & 0.00000 & 0.00000 & 0.80000 & 0.00000 \\
5 & H1 & -12.50000 & 0.00000 & 0.86000 & 0.00000 \\
6 & H1 & -5.00000 & 0.00000 & 0.79937 & 0.00000 \\
7 & H1 & -4.00000 & 0.00000 & 0.23000 & 0.00000 \\
8 & H1 & 4.06289 & 0.00000 & 0.62046 & 0.00000 \\
\midrule
0 & H2 & -0.05089 & 0.00000 & 14.99987 & 2.88027 \\
1 & H2 & 0.00000 & 0.00000 & 0.40000 & 2.88047 \\
2 & H2 & 0.00000 & -0.00075 & 0.40001 & 2.88137 \\
3 & H2 & 19.80580 & 0.00000 & 0.63000 & 0.00000 \\
4 & H2 & 0.00000 & 0.00000 & 0.80000 & 0.00000 \\
5 & H2 & -12.50000 & 0.00000 & 0.86000 & 0.00000 \\
6 & H2 & -5.00000 & 0.00000 & 0.79937 & 0.00000 \\
7 & H2 & -4.00000 & 0.00000 & 0.23000 & 0.00000 \\
8 & H2 & 4.06300 & 0.00000 & 0.63000 & 0.00000 \\
\midrule
0 & H3 & 0.00000 & 0.00000 & 14.95132 & -0.04236 \\
1 & H3 & 0.00000 & 0.00000 & 0.39870 & -0.04236 \\
2 & H3 & 0.00000 & -0.00075 & 0.39870 & -0.04235 \\
3 & H3 & 20.00000 & 0.00000 & 0.62046 & 0.00000 \\
4 & H3 & 0.00000 & 0.00000 & 0.80000 & 0.00000 \\
5 & H3 & -12.50000 & 0.00000 & 0.86000 & 0.00000 \\
6 & H3 & -5.00000 & 0.00000 & 0.79937 & 0.00000 \\
7 & H3 & -4.00000 & 0.00000 & 0.23000 & 0.00000 \\
8 & H3 & 3.81736 & 0.00000 & 0.62046 & 0.00000 \\
\midrule
0 & H4 & -2.49415 & 0.00000 & 14.75867 & 0.83007 \\
1 & H4 & 0.00000 & 0.00000 & 0.39361 & 0.83667 \\
2 & H4 & 0.00000 & 0.00166 & 0.39408 & 0.88505 \\
3 & H4 & 22.14000 & 0.00000 & 0.62046 & 0.00000 \\
4 & H4 & 0.00000 & 0.00000 & 0.80000 & 0.00000 \\
5 & H4 & -12.50000 & 0.00000 & 0.86000 & 0.00000 \\
6 & H4 & -5.00000 & 0.00000 & 0.79937 & 0.00000 \\
7 & H4 & -4.00000 & 0.00000 & 0.23000 & 0.00000 \\
8 & H4 & 4.20000 & 0.00000 & 0.62046 & 0.00000 \\
\bottomrule
\end{tabular}
\end{table}

\begin{table}[H]
\caption{CEDER microgrid measurements for different scenarios.\label{tab:meas}}
\begin{tabular}{ccccccc}
\toprule
\textbf{Bus $i$} & \textbf{Scenario} & \textbf{$P_i$ (kW)} & \textbf{$V_i$ (kV)} & \textbf{Scenario} & \textbf{$P_i$ (kW)} & \textbf{$V_i$ (kV)} \\
\midrule
0 & H1 & 0.000 & 14.790 & H2 & -0.050 &14.990 \\
1 & H1 & 0.000 & 0.399 & H2 & 0.000 & 0.399 \\
2 & H1 & 0.000 & 0.399 & H2 & 0.000 & 0.399 \\
3 & H1 & 19.750 & 0.612 & H2 & 19.800 & 0.611 \\
4 & H1 & 0.000 & 0.799 & H2 & 0.000 & 0.808 \\
5 & H1 & -12.500 & 0.859 & H2 & -12.500 & 0.862 \\
6 & H1 & -5.000 & 0.799 & H2 & -5.000 & 0.799 \\
7 & H1 & -4.000 & 0.230 & H2 & -4.000 & 0.230 \\
8 & H1 & 4.100 & 0.629 & H2 & 4.102 & 0.630 \\
\midrule
0 & H3 & 0.000 & 14.920 & H4 & -2.490 & 14.880 \\
1 & H3 & 0.000 & 0.398 & H4 & 0.000 & 0.399 \\
2 & H3 & 0.000 & 0.398 & H4 & 0.000 & 0.399 \\
3 & H3 & 20.000 & 0.613 & H4 & 22.140 & 0.632 \\
4 & H3 & 0.000 & 0.806 & H4 & 0.000 & 0.805 \\
5 & H3 & -12.500 & 0.860 & H4 & -12.500 & 0.865 \\
6 & H3 & -5.000 & 0.801 & H4 & -5.000 & 0.799 \\
7 & H3 & -4.000 & 0.230 & H4 & -4.000 & 0.230 \\
8 & H3 & 3.800 & 0.628 & H4 & 4.200 & 0.625 \\
\bottomrule
\end{tabular}
\end{table}

\begin{table}[H]
\caption{KPIs results.\label{tab:kpi}}
\begin{tabular}{ccccccccccc}
\toprule
\textbf{Scenario} & \textbf{KPI1 (kWh)}	& \textbf{KPI2 (kgCO\textsubscript{2})} & \textbf{KPI3 (\%)} & \textbf{KPI4 (kW)} \\
\midrule
Baseline & 80,425 & 2,346 & 102.01 & 25 \\
No battery & 80,425 & 2,346 & 102.01 & 0 \\
Battery flexibility & 80,425 & 2,346 & 102.01 & 25 \\
VB flexibility & 80,425 & 2,346 & 102.01 & 25 \\
\bottomrule
\textbf{Scenario} & \textbf{KPI5 (\euro{})} & \textbf{KPI6 (\euro{})} & \textbf{KPI7 (years)} & \textbf{KPI8 (\euro{}/kWh)} \\
\midrule
Baseline & 256,824.77 & 260,635.35 & Never & 145.73 \\
No Battery & 256,824.77 & 204,286.56 & 21 & 113.92 \\
Battery & 301,874.24 & 260,635.35 & 24 & 145.73 \\
VB flexibility & 301,874.24 & 204,286.56 & 17 & 113.92 \\
\bottomrule
\end{tabular}
\end{table}


\end{document}